\documentclass[amsmath,amssymb,prd,showpacs,nofootinbib,floatfix,twocolumn]{revtex4}
\usepackage{epsfig}
\usepackage{xspace}	
\newcommand{\lanl}{Los Alamos National Laboratory, Los Alamos, NM 87545, USA}
\newcommand{\colorado}{University of Colorado, Boulder, CO 80309, USA}
\newcommand{\massinsttech}{Massachusetts Institute of Technology, Cambridge, MA 02139-4307, USA}

\newcommand{\regensburg}{Institut f\"{u}r Theoretische Physik, Universit\"{a}t Regensburg, 93040 Regensburg, Germany}
\newcommand{\buenosaires}{Instituto de F\'{\i}sica de Buenos Aires, CONICET, Departamento de F\'{\i}sica, 
Facultad de Ciencias Exactas y Naturales,\\ Universidad de Buenos Aires, Ciudad Universitaria, Pabell\'on\ 1 (1428) Buenos Aires, Argentina}
\newcommand{\mainz}{Universit\"{a}t Mainz, Institut f\"{u}r Physik, 55099 Mainz, Germany}
\begin{document}
%
\title{Global Analysis of Fragmentation Functions for Eta Mesons}
\author{Christine~A.~Aidala} \email{caidala@bnl.gov} \affiliation{\lanl}
\author{Frank~Ellinghaus} \email{ellingha@uni-mainz.de} \affiliation{\colorado}\affiliation{\mainz}
\author{Rodolfo~Sassot} \email{sassot@df.uba.ar} \affiliation{\buenosaires}
\author{Joseph~P.~Seele} \email{seelej@mit.edu} \affiliation{\colorado}\affiliation{\massinsttech}
\author{Marco~Stratmann} \email{marco@ribf.riken.jp}\thanks{address after Oct.~$1^{\mathrm{st}}$: Physics Department,
Brookhaven National Laboratory, Upton, NY~11973, USA} \affiliation{\regensburg}
\begin{abstract}
Fragmentation functions for eta mesons are extracted 
at next-to-leading order accuracy of QCD in a global analysis of
data taken in electron-positron annihilation and proton-proton scattering experiments.
The obtained parametrization is in good agreement with all data sets analyzed and can be 
utilized, for instance, in future studies of double-spin asymmetries for single-inclusive
eta production.
The Lagrange multiplier technique is used to estimate the uncertainties of the
fragmentation functions and to assess the role of the different data sets
in constraining them.
\end{abstract}
\maketitle
%
\section{Introduction}
%
Fragmentation functions (FFs) are a key ingredient in the perturbative QCD (pQCD) 
description of processes with an observed hadron in the final-state.
Similar to parton distribution functions (PDFs), which account for the 
universal partonic structure of the interacting hadrons, FFs
encode the non-perturbative details of the hadronization process \cite{ref:ffdef}.
When combined with the perturbatively calculable hard scattering cross sections,
FFs extend the ideas of factorization to a much wider class of processes
ranging from hadron production in electron-positron annihilation to
semi-inclusive deep-inelastic scattering (SIDIS) and hadron-hadron collisions \cite{ref:fact}.

Over the last years, our knowledge on FFs has improved dramatically \cite{ref:ff-overview} from 
first rough models of quark and gluon hadronization probabilities \cite{ref:feynman}
to rather precise global analyses at next-to-leading order (NLO) accuracy of QCD,
including estimates of uncertainties \cite{ref:dsspion,ref:dssproton,ref:akk,ref:hirai}.
While the most accurate and clean information used to determine FFs comes from
single-inclusive electron-positron annihilation (SIA) into hadrons,
such data do not allow disentanglement of quark from anti-quark fragmentation
and constrain the gluon fragmentation only weakly through scaling violations
and sub-leading NLO corrections.
Modern global QCD analyses \cite{ref:dsspion,ref:dssproton}
utilize to the extent possible complementary measurements of hadron spectra 
obtained in SIDIS and hadron-hadron collisions to circumvent these shortcomings and to
constrain FFs for all parton flavors individually.

Besides the remarkable success of the pQCD 
approach in describing all the available data simultaneously,
the picture emerging from such comprehensive studies reveals
interesting and sometimes unexpected patterns between the FFs for 
different final-state hadrons.
For instance, the strangeness-to-kaon fragmentation function obtained in Ref.~\cite{ref:dsspion}
is considerably larger than those assumed previously in analyses of SIA data alone
\cite{ref:kretzer}. This has a considerable impact on the extraction of
the amount of strangeness polarization in the nucleon \cite{ref:dssv}
from SIDIS data, which in turn is linked to the fundamental question
of how the spin of the nucleon is composed of intrinsic spins and orbital angular
momenta of quarks and gluons.

Current analyses of FFs comprise pions, kaons, protons
\cite{ref:dsspion,ref:dssproton,ref:akk,ref:hirai}, and lambdas \cite{ref:dsv,ref:akk}
as final-state hadrons.
In this respect, FFs are a much more versatile tool to explore
non-perturbative aspects of QCD than PDFs where studies
are mainly restricted to protons \cite{ref:cteq,ref:otherpdfs}.
In the following, we extend the global QCD analyses of FFs 
at NLO accuracy as described in Refs.~\cite{ref:dsspion,ref:dssproton} 
to eta mesons and estimate the respective uncertainties with the
Lagrange multiplier method \cite{ref:lagrange,ref:dsspion,ref:dssv}.
We obtain a parametrization from experimental data for single-inclusive
eta meson production in SIA at various center-of-mass system (c.m.s.) energies
$\sqrt{S}$ and proton-proton collisions at BNL-RHIC in a wide range
of transverse momenta $p_T$.
We note two earlier determinations of eta FFs in 
Refs.~\cite{ref:greco} and \cite{ref:indumathi} which are based on
normalizations taken from a Monte Carlo event generator and $SU(3)$ model
estimates, respectively. In both cases, parametrizations are not available.

The newly obtained FFs provide fresh insight into the hadronization process
by comparing to FFs for other hadrons.
In particular, the peculiar wave function of the eta,
$|\eta\rangle\simeq |u\bar{u}+d\bar{d}-2s\bar{s}\rangle$, 
with all light quarks and anti-quarks being present,
may reveal new patterns between FFs for different partons and hadrons.
The similar mass range of kaons and etas,
$m_{K^0}\simeq 497.6\,\mathrm{MeV}$ and $m_{\eta}\simeq 547.9\,\mathrm{MeV}$,
respectively, and the presence of strange quarks in both wave functions
makes comparisons between the FFs for these mesons especially relevant.
Of specific interest is also the apparently universal ratio of eta
to neutral pion yields for $p_T\gtrsim 2\,\mathrm{GeV}$ 
in hadron-hadrons collisions across a wide range of c.m.s.~energies,
see, e.g., Ref.~\cite{ref:phenix2006}, and how this is compatible
with the extracted eta and pion FFs.
 
In addition, the availability of eta FFs permits for the first time
NLO pQCD calculations of double-spin asymmetries for single-inclusive 
eta meson production at high $p_T$ which have been 
measured at RHIC \cite{ref:ellinghaus} recently. 
Such calculations are of topical interest for global QCD analyses of
the spin structure of the nucleon \cite{ref:dssv}. 
Finally, the set of eta FFs also provides the baseline
for studies of possible modifications in a nuclear medium \cite{ref:nuclreview,ref:nffs},
for instance, in deuteron-gold collisions at RHIC \cite{ref:phenix2006}.

The remainder of the paper is organized as follows: next, we give a brief outline
of the analysis. In Sec.~\ref{sec:results} we present the results for the eta
FFs, compare to data, and discuss our estimates of uncertainties. We conclude
in Sec.~\ref{sec:conclusions}.

\section{Outline of the Analysis\label{sec:outline}}
%
\subsection{Technical framework and parametrization \label{subsec:outline}}
The pQCD framework at NLO accuracy
for the scale evolution of FFs \cite{ref:evol} 
and single-inclusive hadron production cross sections in
SIA \cite{ref:eenlo} and hadron-hadron collisions \cite{ref:ppnlo} 
has been in place for quite some time and does not need to be repeated here.
Likewise, the global QCD analysis of the eta FFs itself follows closely
the methods outlined in a corresponding fit of pion and kaon FFs
in Ref.~\cite{ref:dsspion}, where all the details can be found. 
As in \cite{ref:dsspion,ref:dssproton} we use the Mellin technique as described in 
\cite{ref:mellin,ref:dssv} to implement all NLO expressions.
Here, we highlight the differences to similar analyses of pion and kaon FFs and 
discuss their consequences for our choice of the functional form 
parameterizing the FFs of the eta meson.

As compared to lighter hadrons, in particular pions, data with identified eta mesons
are less abundant and less precise.
Most noticeable is the lack of any experimental information from SIDIS so far,
which provided the most important constraints on the separation 
of contributions from $u$, $d$, and $s$ (anti-)quarks fragmenting into pions and kaons 
\cite{ref:dsspion}.
Since no flavor-tagged data exist for SIA either, it is inevitable that
a fit for eta FFs has considerably less discriminating power.
Hence, instead of extracting the FFs for the light quarks and anti-quarks 
individually, we parametrize the flavor singlet combination 
at an input scale of $\mu_0=1\,\mathrm{GeV}$,
assuming that all FFs are equal, i.e.,
$D^{\eta}_u=D^{\eta}_{\bar{u}}=D^{\eta}_d=D^{\eta}_{\bar{d}}=
D^{\eta}_s=D^{\eta}_{\bar{s}}$.
We use the same flexible functional form as in Ref.~\cite{ref:dsspion}
with five fit parameters,
\begin{eqnarray}
\label{eq:ansatz}
D_{i}^{\eta}(z,\mu_0)  =  
N_{i} \,z^{\alpha_{i}}(1-z)^{\beta_{i}} [1+\gamma_{i} (1-z)^{\delta_{i}}]\,\,\, \times\,\,\,\,\,\,\, \nonumber\\
\,\,\,\,\,\,\,\,\,\,\,\,\,\,\,\,\,\,\,\,\,
\frac{1}{B[2+\alpha_{i},\beta_{i}+1]+\gamma_i B[2+\alpha_{i},\beta_{i}+\delta_{i}+1]}\;,
\end{eqnarray}
where $z$ is the fraction of the four-momentum of the parton taken by the 
eta meson and $i=u,\bar{u},d,\bar{d},s,\bar{s}$.
$B[a,b]$ denotes the Euler beta function with $a$ and $b$ chosen such that $N_i$
is normalized to the second moment $\int_0^1 zD_i^{\eta}(z,\mu_0)\, dz$
of the FFs.

Although the assumption of equal light quark FFs seems to be rather 
restrictive at first, such an ansatz can be anticipated 
in view of the wave function of the eta meson. 
One might expect a difference between strange and non-strange FFs though
due to the larger mass of strange quarks, i.e., that the hadronization
of $u$ or $d$ quarks is somewhat less likely as they need to pick
up an $s\bar{s}$ pair from the vacuum to form the eta.
Indeed, a ``strangeness suppression" is found for kaon FFs \cite{ref:dsspion}
leading, for instance, to $D_s^{K^{-}}>D_{\bar{u}}^{K^{-}}$. 
In case of the eta wave function one can argue, however, that also a fragmenting $s$ quark
needs to pick up an $s\bar{s}$ pair from the vacuum.
Nevertheless, we have explicitly checked that the introduction
of a second independent parameterization like in (\ref{eq:ansatz}) 
to discriminate between the strange and 
non-strange FFs, does not improve the quality of the fit to the 
currently available data. Clearly, SIDIS data would be
required to further refine our assumptions in the light quark sector
in the future.

The gluon-to-eta fragmentation $D_g^{\eta}$ is mainly constrained by 
data from RHIC rather than scaling violations in SIA. As for pion
and kaon FFs in \cite{ref:dsspion}, we find that a simplified functional
form with $\gamma_g=0$ in Eq.~(\ref{eq:ansatz})
provides enough flexibility to accommodate all data.

Turning to the fragmentation of heavy charm and bottom quarks into eta mesons,
we face the problem that none of the available data sets constraints
their contributions significantly. Here, the lack of any flavor-tagged 
data from SIA hurts most as hadron-hadron cross sections at RHIC energies 
do not receive any noticeable contributions from heavy quark fragmentation.
Introducing independent FFs for charm and bottom at their respective mass
thresholds improves the overall quality of the fit 
but their parameters are essentially unconstrained.
For this reason, we checked that taking the shape of the much better constrained
charm and bottom FFs for pions, kaons, protons, and residual charged hadrons
from \cite{ref:dsspion,ref:dssproton}, but allowing for different normalizations,
leads to fits of comparable quality with only two additional free parameters.

The best fit is obtained for the charm and bottom FFs from an analysis
of residual charged hadrons \cite{ref:dssproton}, 
i.e., hadrons other than pions, kaons, and protons, and hence we use
\begin{eqnarray}
\label{eq:ansatz-hq}
D_{c}^{\eta}(z,m_c) &=& D_{\bar{c}}^{\eta}(z,m_c) = N_c \,D_{c}^{res}(z,m_c)\;, \nonumber \\
D_{b}^{\eta}(z,m_b) &=& D_{\bar{b}}^{\eta}(z,m_b) = N_b \,D_{b}^{res}(z,m_b)\;.
\end{eqnarray}
$N_c$ and $N_b$ denote the normalizations for the charm and bottom fragmentation probabilities
at their respective initial scales, to be constrained by the fit to data.
The parameters specifying the $D_{c,b}^{res}$ can be found in Tab.~III of Ref.~\cite{ref:dssproton}.
The FFs in Eq.~(\ref{eq:ansatz-hq}) are included discontinuously as massless
partons in the scale evolution of the FFs above their $\overline{\mathrm{MS}}$
thresholds $\mu=m_{c,b}$ with $m_{c}=1.43\,\mathrm{GeV}$ and $m_{b}=4.3\,\mathrm{GeV}$ 
denoting the mass of the charm and bottom quark, respectively.

In total, the parameters introduced in Eqs.~(\ref{eq:ansatz}) and (\ref{eq:ansatz-hq})
to describe the FFs of quarks and gluons into eta mesons add up to 10.
They are determined by a standard $\chi^2$ minimization for 
$N=140$ data points, where
\begin{equation}
\label{eq:chi2}
\chi^2=\sum_{j=1}^N \frac{(T_j-E_j)^2}{\delta E_j^2}\;.
\end{equation}
$E_j$ represents the experimentally measured value of a given observable, 
$\delta E_j$ its associated uncertainty, and $T_j$ is the 
corresponding theoretical estimate calculated at NLO accuracy for a given
set of parameters in Eqs.~(\ref{eq:ansatz}) and (\ref{eq:ansatz-hq}).
For the experimental uncertainties $\delta E_i$ we take 
the statistical and systematic errors in quadrature for the time being.

\subsection{Data sets included in the fit\label{subsec:data}}
%
A total of 15 data sets is included in our analysis.
We use all SIA data with $\sqrt{S}>10\,\mathrm{GeV}$:
HRS \cite{ref:hrs} and MARK II \cite{ref:mark2} 
at $\sqrt{S}=29\,\mathrm{GeV}$,
JADE \cite{ref:jade1,ref:jade2} and CELLO \cite{ref:cello} 
at $\sqrt{S}=34-35\,\mathrm{GeV}$, and 
ALEPH \cite{ref:aleph1,ref:aleph2,ref:aleph3}, 
L3 \cite{ref:l31,ref:l32}, and OPAL \cite{ref:opal} 
at $\sqrt{S} = M_Z = 91.2\,\mathrm{GeV}$.
Preliminary results from BABAR \cite{ref:babar} at 
$\sqrt{S}=10.54\,\mathrm{GeV}$
are also taken into account.  

The availability of $e^+e^-$ data in approximately 
three different energy regions of $\sqrt{S} \simeq 10,$
30, and $90\,\mathrm{GeV}$ helps to constrain the gluon
fragmentation function from scaling violations.
Also, the appropriate electroweak charges in the
inclusive process $e^+e^-\to (\gamma,Z)\rightarrow \eta X$ 
vary with energy, see, e.g.,
App.~A of Ref.~\cite{ref:dsv} for details, 
and hence control which combinations of quark FFs are probed. 
Only the CERN-LEP data taken on the $Z$ resonance 
receive significant contributions from charm and bottom FFs.

Given that the range of applicability for FFs is limited to 
medium-to-large values of the energy fraction $z$, 
as discussed, e.g., in Ref.~\cite{ref:dsspion}, 
data points with $z<0.1$ are excluded from the fit.  
Whenever the data set is expressed in terms of the scaled
three-momentum of the eta meson, i.e., $x_p\equiv 2p_{\eta}/\sqrt{S}$, 
we convert it to the usual scaling variable $z=x_p/\beta$, where
$\beta=p_{\eta}/E_{\eta}=\sqrt{1-m_{\eta}^2/E_{\eta}^2}$.
In addition to the cut $z>0.1$, we also impose that
$\beta>0.9$ in order to avoid kinematic regions where 
mass effects become increasingly relevant.
The cut on $\beta$ mainly affects the data at 
low $z$ from BABAR \cite{ref:babar}.

In case of single-inclusive eta meson production in
hadron-hadron collisions, we include data sets from PHENIX
at $\sqrt{S}=200\,\mathrm{GeV}$ at mid-rapidity
\cite{ref:phenix2006,ref:phenix-run6}
in our global analysis. 
The overall scale uncertainty of $9.7\%$ in the PHENIX measurement
is not included in $\delta E_j$ in Eq.~(\ref{eq:chi2}).
All data points have a transverse
momentum $p_T$ of at least $2\,\mathrm{GeV}$.
As we shall demonstrate below, these data provide an
invaluable constraint on the quark and gluon-to-eta fragmentation
probabilities. 
In general, hadron collision data
probe FFs at fairly large momentum fractions $z\gtrsim 0.5$,
see, e.g., Fig.~6 in Ref.~\cite{ref:lhcpaper}, complementing the
information available from SIA. 
The large range of $p_T$ values covered by the recent PHENIX
data \cite{ref:phenix-run6}, $2\le p_T\le 20\,\mathrm{GeV}$,
also helps to constrain FFs through scaling violations.

As in other analyses of FFs \cite{ref:dsspion,ref:dssproton}
we do not include eta meson production data from 
hadron-hadron collision experiments at much lower c.m.s.~energies, 
like Fermilab-E706 \cite{ref:e706}. It is known that
theoretical calculations at NLO accuracy do not reproduce
such data very well without invoking resummations of
threshold logarithms to all orders in pQCD \cite{ref:resum}.

\section{Results\label{sec:results}}
%
In this Section we discuss in detail the results of our
global analysis of FFs for eta mesons
at NLO accuracy of QCD.
First, we shall present the parameters of the optimum
fits describing the $D_i^{\eta}$ at the input scale.
Next, we compare our fits to the data used in the
analysis and give $\chi^2$ values for each individual set
of data.
Finally, we estimate the uncertainties in the extraction
of the $D_i^{\eta}$ using the Lagrange multiplier technique 
and discuss the role of the different 
data sets in constraining the FFs.

\subsection{Optimum fit to data \label{subsec:fit}}
%
In Tab.~\ref{tab:para} we list the set of parameters specifying the
optimum fit of eta FFs at NLO accuracy in
Eqs.~(\ref{eq:ansatz}) and (\ref{eq:ansatz-hq}) at our
input scale $\mu_0=1\,\mathrm{GeV}$ for the light 
quark flavors and the gluon. Charm and bottom FFs
are included at their mass threshold
$\mu_0=m_c$ and $\mu_0=m_b$, respectively \cite{ref:fortran}.

The data sets included in our global analysis, as discussed in
Sec.~\ref{subsec:data}, and the individual $\chi^2$ values 
are presented in Tab.~\ref{tab:data}. 
%
%
\begin{table}[bh!]
\caption{\label{tab:para}Parameters describing the NLO
FFs for eta mesons, $D_i^{\eta}(z,\mu_0)$,
in Eqs.~(\ref{eq:ansatz}) and (\ref{eq:ansatz-hq})
at the input scale $\mu_0=1\,\mathrm{GeV}$.
Inputs for the charm and bottom FFs refer to
$\mu_0=m_c$ and $\mu_0=m_b$, respectively.}
\begin{ruledtabular}
\begin{tabular}{cccccc}
Flavor $i$ &$N_i$ & $\alpha_i$ & $\beta_i$& $\gamma_i$& $\delta_i$ \\
\hline
$u,\bar{u},d,\bar{d},s,\bar{s}$& 0.038 & 1.372 & 1.487 & 2000.0 & 34.03\\
$g$    & 0.070 & 10.00 & 9.260 & 0 & 0\\
$c,\bar{c}$& 1.051  & - & - & - & - \\
$b,\bar{b}$& 0.664 & - & - & - & - \\
\end{tabular}
\end{ruledtabular}
\end{table}
%
%
%
\begin{figure*}[ht!]
\begin{center}
\vspace*{-0.6cm}
\epsfig{figure=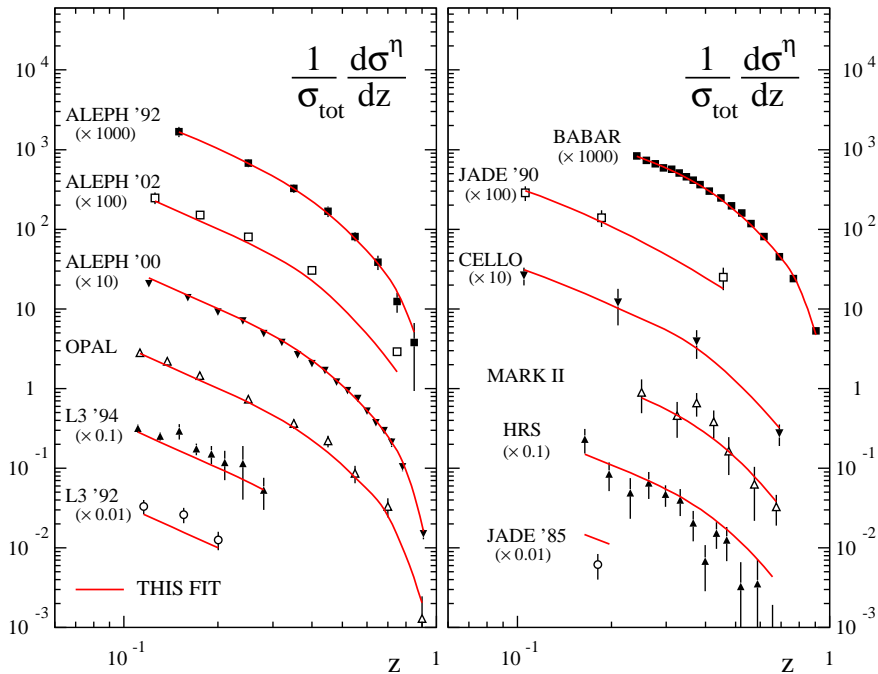,width=0.69\textwidth}
\end{center}
\vspace*{-0.7cm}
\caption{Comparison of our NLO results 
with the data sets for inclusive eta meson production in SIA used in the fit;
see Tab.~\ref{tab:data}.
\label{fig:sia-eta}}
\vspace*{-0.5cm}
\end{figure*}
We note that the quoted number of points and $\chi^2$ values
are based only on fitted data, i.e., $z>0.1$ and $\beta>0.9$ in SIA. 

As can be seen, for most sets of data their partial contribution to
the $\chi^2$ of the fit is typically of the order of the number of data points
or even smaller.
The most notable exceptions are the HRS \cite{ref:hrs} and ALEPH~'02 \cite{ref:aleph3}
data, where a relatively small number of points have a significant $\chi^2$, which
in turn leads to total $\chi^2$ per degree of freedom (d.o.f.) of about 1.6 for the fit.
We have checked that these more problematic sets of data could be removed from the fit 
without reducing its constraining power or changing the obtained $D_i^{\eta}$
significantly.
The resulting, fairly large $\chi^2/d.o.f.$ due to a few isolated data points
is a common characteristic of all
extractions of FFs made so far \cite{ref:dsspion,ref:dssproton,ref:akk,ref:hirai,ref:kretzer}
for other hadron species. 

The overall excellent agreement of our fit with experimental results
for inclusive eta meson production in SIA and the tension with 
the HRS and ALEPH~'02 data is also illustrated in Fig.~\ref{fig:sia-eta}.
It is worth pointing out that both ALEPH~'00 \cite{ref:aleph2} and BABAR 
\cite{ref:babar} data are well reproduced for all momentum fractions $z$
in spite of being at opposite ends of the c.m.s.~energy range 
covered by experiments.
%
%
\begin{table}[bh!]
\caption{\label{tab:data}Data used in the global analysis
of eta FFs, the individual $\chi^2$ values for
each set,  and the total $\chi^2$ of the fit.}
\begin{ruledtabular}
\begin{tabular}{lrr}
Experiment & data points & $\chi^2$ \\
                 & fitted     &         \\\hline
BABAR \cite{ref:babar}    &  18 & 8.1 \\
HRS \cite{ref:hrs}        &  13 & 51.6 \\
MARK II \cite{ref:mark2}  &   7 & 3.8 \\
JADE '85 \cite{ref:jade1} &   1 & 9.6\\
JADE '90 \cite{ref:jade2} &   3 & 1.2\\
CELLO \cite{ref:cello}    &   4 & 1.1 \\
ALEPH '92 \cite{ref:aleph1}   & 8 & 2.0 \\
ALEPH '00 \cite{ref:aleph2}  & 18 & 22.0 \\
ALEPH '02 \cite{ref:aleph3}   & 5 & 61.6 \\
L3 '92 \cite{ref:l31}   & 3 &  5.1 \\
L3 '94 \cite{ref:l32}   & 8 &  10.5\\
OPAL \cite{ref:opal}    & 9 &  9.0\\
PHENIX $2 \gamma$ \cite{ref:phenix2006}   & 12 & 4.1 \\
PHENIX $3 \pi$ \cite{ref:phenix2006}      &  6 & 2.9 \\
PHENIX '06 \cite{ref:phenix-run6}         & 25 & 13.3 \\
\hline
{\bf TOTAL:} &  140 &  205.9 \\
\end{tabular}
\end{ruledtabular}
\end{table}

%
%
\begin{figure}[t]
\begin{center}
\vspace*{-0.6cm}
\epsfig{figure=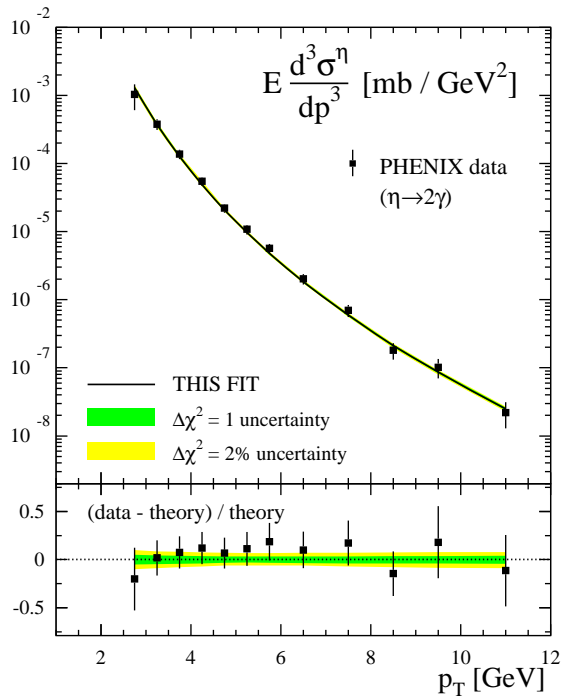,width=0.48\textwidth}
\end{center}
\vspace*{-0.7cm}
\caption{Upper panel: comparison of our NLO result for single-inclusive eta
production in $pp$ collisions at $\sqrt{S}=200\,\mathrm{GeV}$
with PHENIX data where the eta is identified in the decay $\eta\to2\gamma$
\cite{ref:phenix2006}. Lower panel:
the ratio (data-theory)/theory. The shaded bands correspond to alternative
fits consistent with an increase of $\Delta \chi^2=1$
or $\Delta \chi^2=2\%$ in the total $\chi^2$ of the best fit (see text).
\label{fig:hadronic2g}}
\vspace*{-0.5cm}
\end{figure}
%
%
\begin{figure}
\begin{center}
\vspace*{-0.6cm}
\epsfig{figure=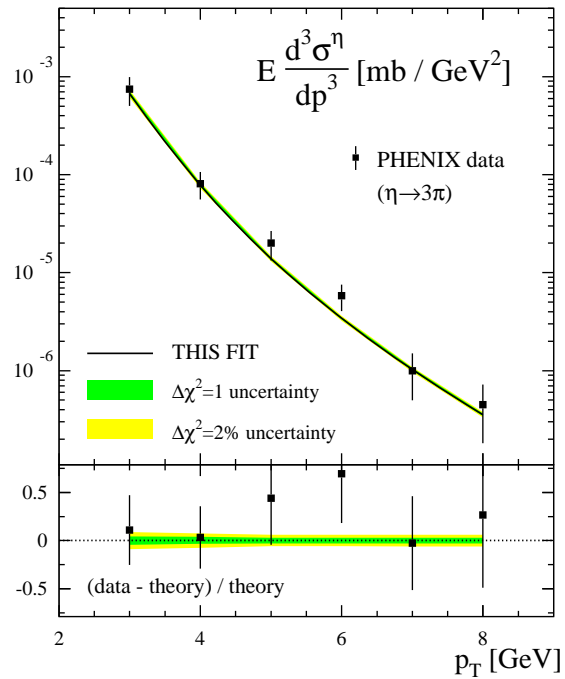,width=0.48\textwidth}
\end{center}
\vspace*{-0.7cm}
\caption{Same as in Fig.~\ref{fig:hadronic2g} but now
for the data where the eta is identified in the decay $\eta\to 3\pi$
\cite{ref:phenix2006}.
\label{fig:hadronic3pi}}
\vspace*{-0.5cm}
\end{figure}
%
%
\begin{figure}
\begin{center}
\vspace*{-0.3cm}
\epsfig{figure=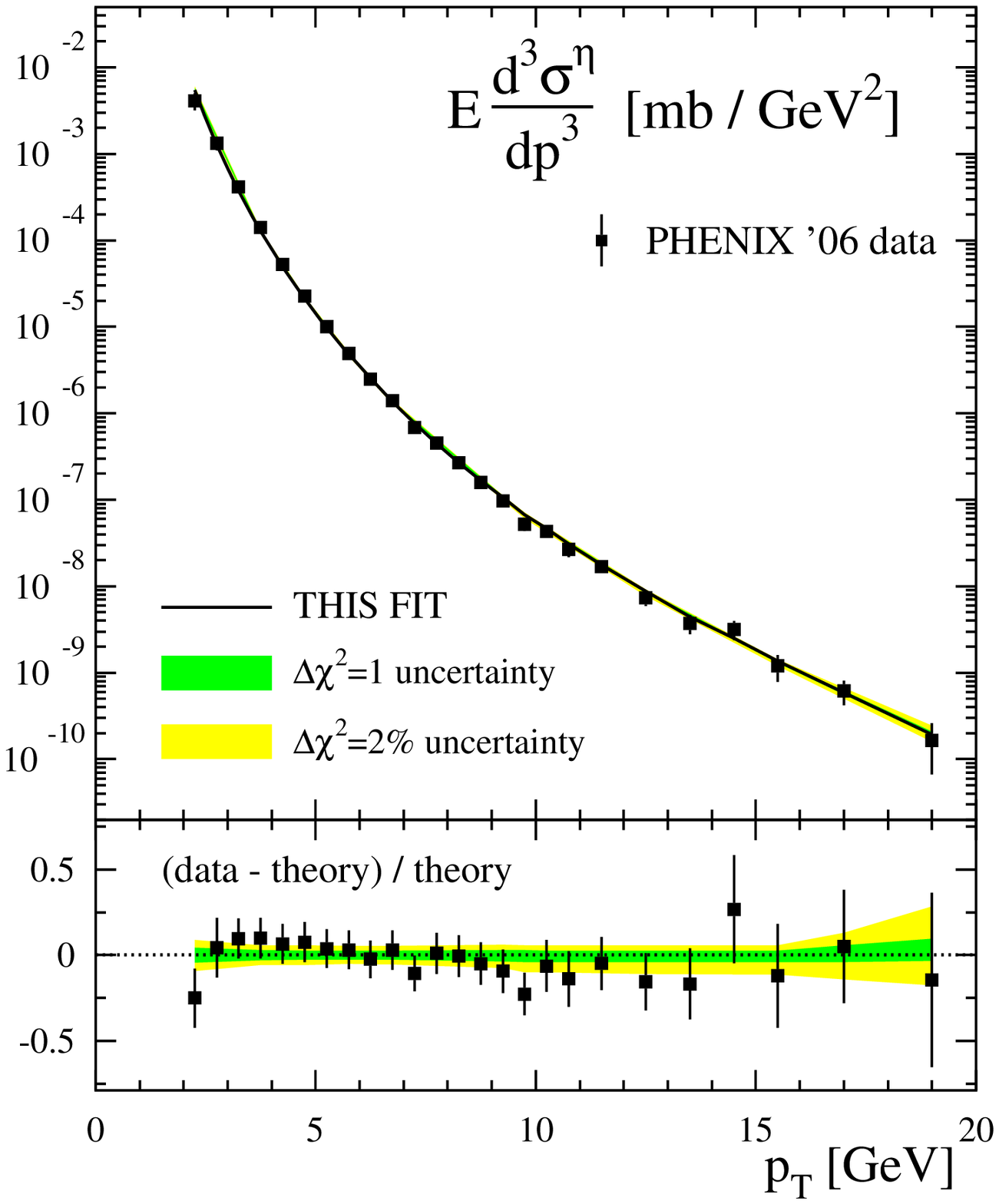,width=0.48\textwidth}
\end{center}
\vspace*{-0.7cm}
\caption{Same as in Fig.~\ref{fig:hadronic2g} but now
for the latest PHENIX data \cite{ref:phenix-run6}.
\label{fig:hadronic06}}
\vspace*{-0.5cm}
\end{figure}
Our fit compares very well with all data on high-$p_T$ eta meson production
in proton-proton collisions from RHIC \cite{ref:phenix2006,ref:phenix-run6}.
The latest set of PHENIX data \cite{ref:phenix-run6} significantly extends
the range in $p_T$ at much reduced uncertainties and provides
stringent constraints on the FFs as we shall demonstrate below.
The normalization and trend of the data is nicely reproduced over a wide
kinematical range as can be inferred from Figs.~\ref{fig:hadronic2g}-\ref{fig:hadronic06}.
In each case, the invariant cross section for $pp\rightarrow \eta X$
at $\sqrt{S}=200\,\mathrm{GeV}$ is computed at NLO accuracy,
averaged over the pseudorapidity range of PHENIX, $|\eta|\le 0.35$,
and using the NLO set of PDFs from CTEQ \cite{ref:cteq} along with the
corresponding value of $\alpha_s$. Throughout our analysis we choose the
transverse momentum of the produced eta as both the factorization
and the renormalization scale, i.e., $\mu_f=\mu_r=p_T$.

Since the cross sections drop over several orders of magnitude
in the given range of $p_T$,
we show also the ratio (data-theory)/theory 
in the lower panels of Figs.~\ref{fig:hadronic2g}-\ref{fig:hadronic06}
to facilitate the comparison between data and our fit.
One notices the trend of the theoretical estimates to overshoot the data
near the lowest values of transverse momenta, $p_T\simeq 2\,\mathrm{GeV}$
which indicates that the factorized pQCD approach starts to fail.
Compared to pion production at central pseudorapidities, see Fig.~6 in
Ref.~\cite{ref:dsspion}, the breakdown of pQCD sets in at somewhat
higher $p_T$ as is expected due to the larger mass of the eta meson.

The shaded bands in Figs.~\ref{fig:hadronic2g}-\ref{fig:hadronic06}
are obtained with the Lagrange multiplier method,
see Sec.~\ref{subsec:uncert} below,
applied to each data point. They correspond to the maximum variation
of the invariant cross section computed with alternative
sets of eta FFs consistent with an increase of $\Delta \chi^2=1$
or $\Delta \chi^2=2\%$ in the total $\chi^2$ of the best global fit
to all SIA and $pp$ data.

%
%
\begin{figure*}
\begin{center}
\vspace*{-0.6cm}
\epsfig{figure=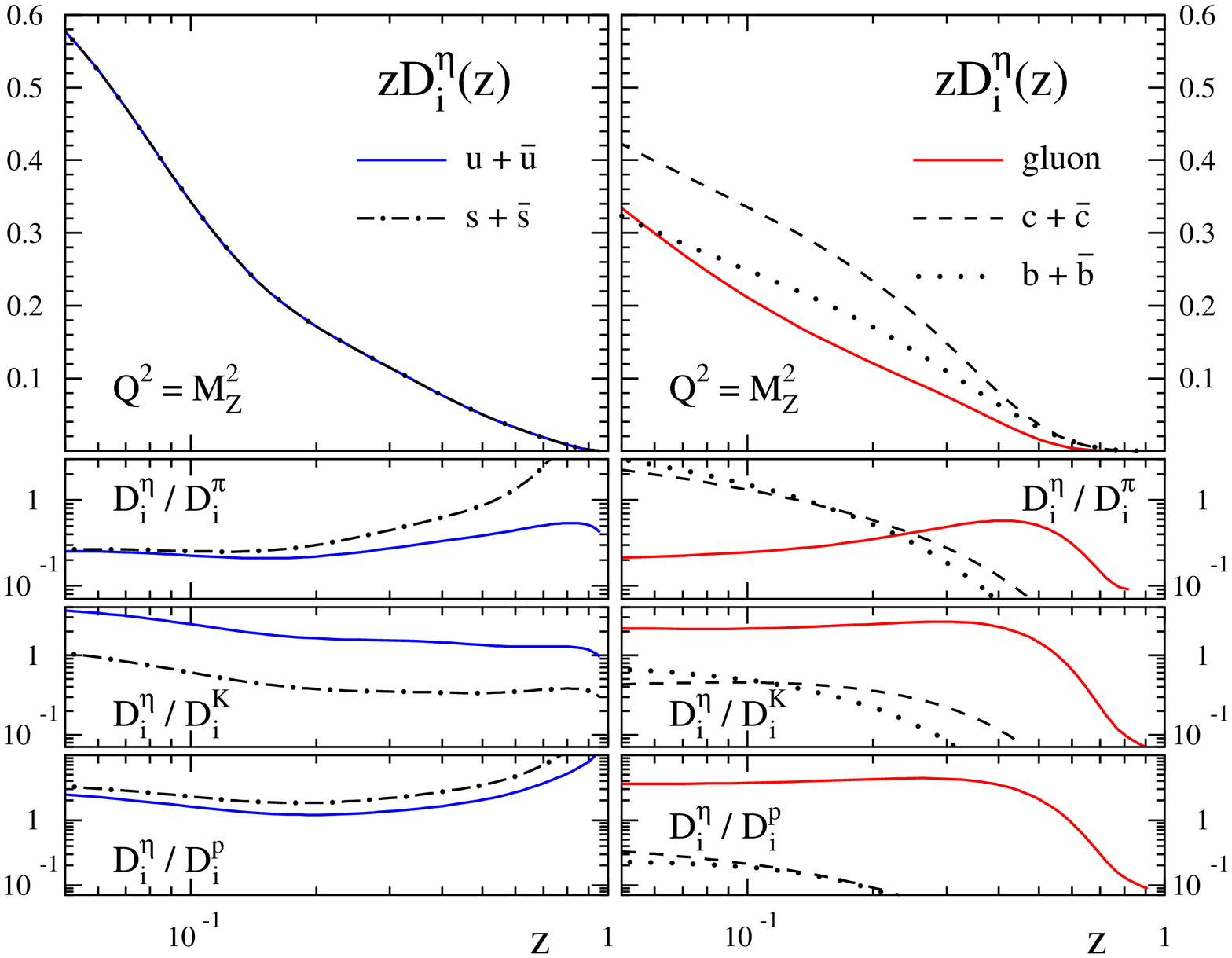,width=0.8\textwidth}
\end{center}
\vspace*{-0.7cm}
\caption{Upper panels: individual FFs for eta mesons
$zD_i^{\eta}(z,Q^2)$ at $Q^2=M_Z^2$
for $i=u+\bar{u}$, $s+\bar{s}$, $g$, $c + \bar{c}$, and $b + \bar{b}$.
Lower three rows of panels: ratios of our eta FFs to the ones for pions, kaons, and protons
as obtained in the DSS analysis \cite{ref:dsspion,ref:dssproton}.
\label{fig:eta-ff-comp}}
\vspace*{-0.5cm}
\end{figure*}
In addition to the experimental uncertainties propagated to the
extracted $D_i^{\eta}$, a large theoretical ambiguity is 
associated with the choice of the factorization and renormalization
scales used in the calculation of the $pp\to \eta X$ cross sections.
These errors are much more sizable than experimental ones and
very similar to those estimated for $pp\to\pi X$ in Fig.~6 of 
Ref.~\cite{ref:dsspion}. 
As in the DSS analysis for pion and kaon FFs \cite{ref:dsspion} the
choice $\mu_f=\mu_r=p_T$ and $\mu_f=\mu_r=S$ in $pp$ collisions
and SIA, respectively, leads to a nice global description of 
all data sets with a common universal set of eta FFs.

Next, we shall present an overview of the obtained FFs $D_i^{\eta}(z,Q)$
for different parton flavors $i$ and compare them to FFs for other hadrons.
The upper row of panels in Fig.~\ref{fig:eta-ff-comp} shows the 
dependence of the FFs on the energy fraction $z$ taken by the eta meson
at a scale $Q$ equal to the mass of the $Z$ boson, i.e., $Q =M_Z$.
Recall that at our input scale $Q=\mu_0=1\,\mathrm{GeV}$ we assume
that $D^{\eta}_u=D^{\eta}_{\bar{u}}=D^{\eta}_d=D^{\eta}_{\bar{d}}=
D^{\eta}_s=D^{\eta}_{\bar{s}}$, which is preserved under scale evolution.
At such a large scale $Q=M_Z$ the heavy quark FFs are of similar size,
which is not too surprising as mass effects are negligible, i.e.,
$m_{c,b}\ll M_Z$. 
The gluon-to-eta fragmentation function $D_g^{\eta}$ is slightly smaller
but rises towards smaller values of $z$.
Overall both the shape and the hierarchy between the different FFs
$D_i^{\eta}$ is similar to those found, for instance, for pions; see
Fig.~18 in \cite{ref:dsspion}, with the exception of the ``unfavored"
strangeness-to-pion fragmentation function which is suppressed.
In order to make the comparison to FFs for other hadrons more explicit,
we show in the lower three rows of Fig.~\ref{fig:eta-ff-comp} the
ratios of the obtained $D_i^{\eta}(z,M_z)$ to the FFs for pions,
kaons, protons from the DSS analysis \cite{ref:dsspion,ref:dssproton}.

The eta and pion production yields are known to be 
consistent with a constant ratio of about a half
in a wide range of c.m.s.~energies in hadronic collisions
for $p_T\gtrsim 2\,\mathrm{GeV}$, but the ratio varies from
approximately 0.2 at $z\simeq 0.1$ to about 0.5 for $z\gtrsim 0.4$ 
in SIA \cite{ref:phenix2006}. 
It is interesting to see how these findings are reflected in
the ratios of the eta and neutral pion FFs for the individual parton flavors.
We find that $D^{\eta}_{u+\bar{u}}/D^{\pi^{0}}_{u+\bar{u}}$ follows closely the
trend of the SIA data as is expected since gluon fragmentation enters only at NLO in
the cross section calculations.
For strangeness the rate of eta to pion FFs increases towards larger $z$ because of
the absence of strange quarks in the pion wave functions.

Inclusive hadron production at small-to-medium values of $p_T$ is known to be
dominated by gluon fragmentation at relatively large values of
momentum fraction $z$ \cite{ref:dsspion,ref:lhcpaper} largely independent of
the c.m.s.~energy $\sqrt{S}$. 
In the relevant range of $z$, $0.4\lesssim z \lesssim 0.6$, the ratio
$D_g^{\eta}/D_g^{\pi^{0}}$ resembles the constant ratio of roughly 0.5
found in the eta-to-pion production yields.
Both at larger and smaller values of $z$ the $D^{\eta}_g$ is suppressed with respect
to $D_g^{\pi^{0}}$.
In general, one should keep in mind that FFs always appear in complicated convolution
integrals in theoretical cross section calculations \cite{ref:eenlo,ref:ppnlo}
which complicates any comparison of cross section and fragmentation function 
ratios for different hadrons.

The comparison to the DSS kaon FFs \cite{ref:dsspion} is shown in the panels in third row 
of Fig.~\ref{fig:eta-ff-comp}. Most remarkable is the ratio of the gluon FFs,
which is approximately constant, $D_g^{\eta}/D_g^K \simeq 2$, 
over a wide region in $z$ but drops below one for $z\gtrsim 0.6$
At large $z$, $D^{\eta}_{u+\bar{u}}$
tends to be almost identical to $D^{K}_{u+\bar{u}}$, while $D^{\eta}_{s+\bar{s}}$ resembles
$D^{K}_{s+\bar{s}}$ only at low $z$.
The latter result might be understood due to the absence of strangeness suppression
for $D^{K}_{s+\bar{s}}$, whereas a fragmenting $s$ quark needs to
pick up an $\bar{s}$ quark from the vacuum to form the eta meson.
It should be noted, however, that kaon FFs have considerably larger uncertainties than
pion FFs \cite{ref:dsspion} which makes the comparisons less conclusive.

This is even more true for the proton FFs \cite{ref:dssproton}.
Nevertheless, it is interesting to compare our $D_i^{\eta}$ to those
for protons which is done in the lower panels of Fig.~\ref{fig:eta-ff-comp}.
As for kaons, we observe a rather flat behavior of the ratio $D_g^{\eta}/D_g^{p}$,
which drops below one at larger values of $z$.
The corresponding rates for light quark FFs show the opposite trend and rise towards
$z\to 1$.

Regarding the relative sizes of the fragmentation probabilities 
for light quarks and gluons into the different hadron species,
we find that eta FFs are suppressed w.r.t.~pion FFs (except for strangeness), 
are roughly similar to those for kaons, and larger than the proton FFs. 
This can be qualitatively understood from the hierarchy of the respective hadron masses.
For $z\gtrsim 0.6$, the lack of decisive constraints from data prevents one from
drawing any conclusions in this kinematic region. 

As we have already discussed in Sec.~\ref{subsec:outline}, due to the lack of any flavor tagged 
SIA data sensitive
to the hadronization of charm and bottom quarks into eta mesons,
we adopted the same functional form as for the fragmentation into
residual charged hadrons \cite{ref:dssproton}, 
i.e., hadrons other than pions, kaons, and protons.
The fit favors a charm fragmentation almost identical to that for the residual 
hadrons  ($N_c=1.058$) and a somewhat reduced distribution for bottom fragmentation
($N_b=0.664$). At variance to what is found for light quarks and gluons,
after evolution, $D_{c+\bar{c}}^{\eta}$ and $D_{b+\bar{b}}^{\eta}$
differ significantly in size and shape from their counterparts for
pions, kaons, and protons as can be also inferred from Fig.~\ref{fig:eta-ff-comp}.
Future data are clearly needed here for any meaningful comparison.

\subsection{Estimates of uncertainties \label{subsec:uncert}}
%
Given the relatively small number of data points available for the determination
of the $D_i^{\eta}$ as compared to global fits of pion, kaon, and
proton FFs \cite{ref:dsspion,ref:dssproton}, we refrain from performing
a full-fledged error analysis. 
However, in order to get some idea of the uncertainties 
of the $D_i^{\eta}$ associated with experimental errors,
how they propagate into observables, and
the role of the different data sets in constraining the $D_i^{\eta}$, 
we perform a brief study based on Lagrange multipliers
\cite{ref:lagrange,ref:dsspion,ref:dssv}.
%
%
\begin{figure*}
\begin{center}
\vspace*{-0.6cm}
\epsfig{figure=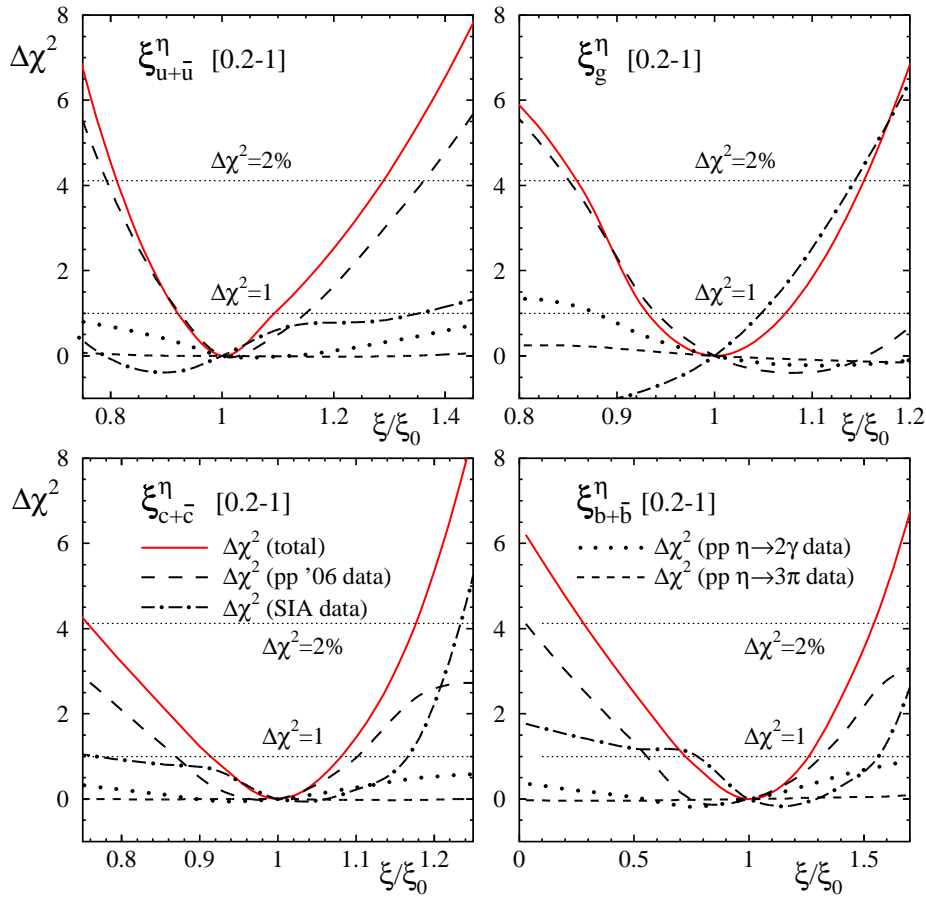,width=0.75\textwidth}
\end{center}
\vspace*{-0.7cm}
\caption{Profiles of $\chi^2$ for the NLO eta fragmentation fit
as a function of the truncated second moments
$\xi^{\eta}_i(z_{\min}=0.2,Q=5\,\text{GeV})$ for different flavors
(solid lines). In each case, the moments are
normalized to the value $\xi_0$ they take in the best fit
to data. The other lines indicate the partial contributions to $\Delta \chi^2$
of the individual data sets used in the fit.
\label{fig:profiles-ffs}}
\vspace*{-0.5cm}
\end{figure*}

This method relates the range of variation of a physical
observable ${\cal{O}}$ dependent on FFs to the variation
in the $\chi^2$ function used to judge the goodness of the fit.
To this end, one minimizes the function
\begin{equation}
\label{eq:lm}
\Phi(\lambda,\{a_i\}) = \chi^2(\{a_i\}) + \lambda\, {\cal{O}}(\{a_i\})
\end{equation}
with respect to the set of parameters $\{a_i\}$ describing the FFs
in Eqs.~(\ref{eq:ansatz}) and (\ref{eq:ansatz-hq})
for fixed values of $\lambda$. Each of the Lagrange multipliers
$\lambda$ is related to an observable ${\cal{O}}(\{a_i\})$, and
the choice $\lambda=0$ corresponds to the optimum global fit.
From a series of fits for different values of $\lambda$ one can map
out the $\chi^2$ profile for any observable ${\cal{O}}(\{a_i\})$
free of the assumptions made in the traditional 
Hessian approach \cite{ref:hessian}.

As a first example and following the DSS analyses \cite{ref:dsspion,ref:dssproton}, 
we discuss the range of variation of the truncated second moments of the 
eta FFs,
\begin{equation}
\label{eq:truncmom}
\xi^{\eta}_i(z_{\min},Q) \equiv \int_{z_{\min}}^1 z D_i^{\eta}(z,Q)\, dz, 
\end{equation}
for $z_{\min}=0.2$ and $Q=5\,\mathrm{GeV}$ around the values obtained
in the optimum fit to data, $\xi^{\eta}_{i\,0}$.
In a LO approximation, the second moments $\int_0^1 zD_i^{\eta}(z,Q)dz$
represent the energy fraction of the parent parton of flavor $i$ 
taken by the eta meson at a scale $Q$.
The truncated moments in Eq.~(\ref{eq:truncmom}) discard the
low-$z$ contributions, which are not constrained by data and, more
importantly, where the framework of FFs does not apply.
In general, FFs enter calculations of cross sections as convolutions
over a wide range of $z$, and, consequently, the $\xi^{\eta}_i(z_{\min},Q)$
give a first, rough idea of how uncertainties in the FFs will propagate 
to observables.

The solid lines in Fig.~\ref{fig:profiles-ffs}
show the $\xi^{\eta}_i(z_{\min},Q)$ defined in Eq.~(\ref{eq:truncmom})
for $i=u+\bar{u}$, $g$, $c+\bar{c}$, and $b+\bar{b}$ 
against the corresponding increase $\Delta \chi^2$ in the total $\chi^2$
of the fit. 
The two horizontal lines indicate a $\Delta \chi^2$ of one unit and
an increase by $2\%$ which amounts to about 4 units in $\chi^2$, 
see Tab.~\ref{tab:data}.
The latter $\Delta \chi^2$ should give a more
faithful estimate of the relevant uncertainties in global QCD analyses
\cite{ref:dsspion,ref:dssproton,ref:cteq,ref:dssv}
than an increase by one unit.

As can be seen, the truncated moment $\xi^{\eta}_{u+\overline{u}}$,
associated with light quark FFs $D^{\eta}_{u+\overline{u}}=
D^{\eta}_{d+\overline{d}}=D^{\eta}_{s+\overline{s}}$, is constrained within
a range of variation of approximately $^{+30\%}_{-20\%}$ around
the value computed with the best fit, assuming a conservative increase
in $\chi^2$ by $2\%$. 
The estimated uncertainties are considerably larger than
the corresponding ones found for pion and kaon FFs, which are
typically of the order of $\pm$3\% and $\pm$10\% for the light quark flavors
\cite{ref:dsspion}, respectively, 
but closer to the $\pm 20\%$ observed for proton 
and anti-proton FFs \cite{ref:dssproton}.
For the truncated moment $\xi_g^{\eta}$ of gluons shown in the upper right 
panel of Fig.~\ref{fig:profiles-ffs}, the range of uncertainty is
slightly smaller than one found for light quarks and amounts to 
about $\pm 15\%$. The allowed variations are
larger for charm and bottom FFs as can be inferred from the lower row of plots 
in Fig.~\ref{fig:profiles-ffs}.

Apart from larger experimental uncertainties and the much smaller
amount of SIA data for identified eta mesons, the lack of any 
information from SIDIS is particularly responsible for the large range of variations
found for the light quarks in Fig.~\ref{fig:profiles-ffs}. We recall that the missing SIDIS
data for produced eta mesons also forced us to assume that all light quark FFs 
are the same in Eq.~(\ref{eq:ansatz}).
The additional ambiguities due to this assumption are not reflected in the
$\chi^2$ profiles shown in Fig.~\ref{fig:profiles-ffs}.
The FFs for charm and bottom quarks into eta mesons suffer most from 
the lack of flavor tagged data in SIA.

To further illuminate the role of the different data sets in constraining the
$D_{i}^{\eta}$ we give also the partial contributions to $\Delta \chi^2$ of
the individual data sets from $pp$ collisions and the combined SIA data
in all panels of Fig.~\ref{fig:profiles-ffs}.
Surprisingly, the light quark FFs are constrained best by the PHENIX $pp$ data
from run~'06 and not by SIA data.
SIA data alone would prefer a smaller value for $\xi^{\eta}_{u+\bar{u}}$
by about $10\%$,
strongly correlated to larger moments for charm and bottom fragmentation,
but the minimum in the $\chi^2$ profile is much less pronounced and very shallow,
resulting in rather sizable uncertainties.

%
%
\begin{figure}[bth!]
\begin{center}
\vspace*{-0.6cm}
\epsfig{figure=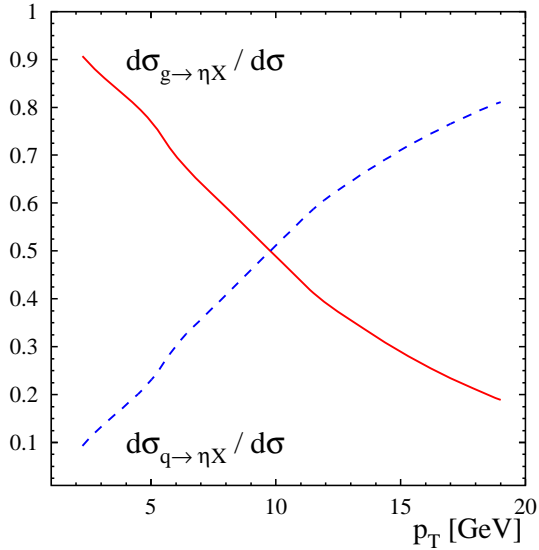,width=0.48\textwidth}
\end{center}
\vspace*{-0.7cm}
\caption{Relative fractions of quarks and gluons fragmenting into observed
eta meson in $pp$ collisions at $\sqrt{S}=200\,\mathrm{GeV}$ and
PHENIX kinematics \cite{ref:phenix-run6}.
\label{fig:fractions}}
\vspace*{-0.5cm}
\end{figure}
%
%
%
\begin{figure*}[hbt!]
\begin{center}
\vspace*{-0.6cm}
\epsfig{figure=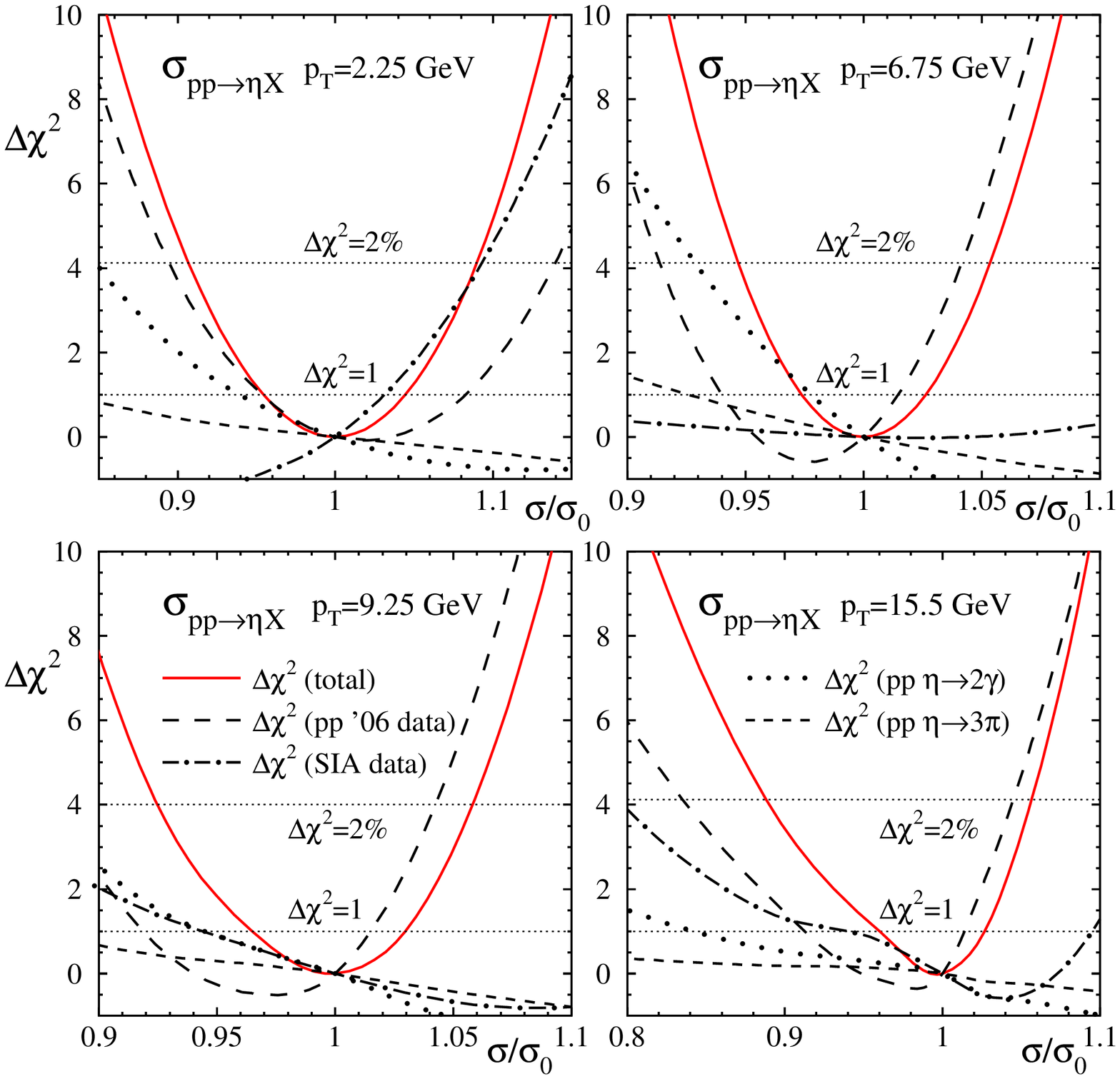,width=0.75\textwidth}
\end{center}
\vspace*{-0.7cm}
\caption{As in Fig.~\ref{fig:profiles-ffs} but now as a function of the
cross section for $pp \rightarrow \eta\,X$ at different values of $p_T$.
The variations of the cross sections are normalized to the value
$\sigma_{0}$ obtained in the optimum fit to data.
\label{fig:profiles-pp}}
\end{figure*}
This unexpected result is most likely due to the fact that the SIA data
from LEP experiments constrain mainly the flavor singlet combination, i.e., the sum of all
quark flavors, including charm and bottom. Since there are no flavor tagged
data available from SIA for eta mesons, the separation into contributions from light and 
heavy quark FFs is largely unconstrained by SIA data. Only the fairly precise data from BABAR
at $\sqrt{S}\simeq 10\,\mathrm{GeV}$ provide some guidance as they constrain a
different combination of the light $u$, $d$, and $s$ quark FFs weighted by the
respective electric charges. Altogether, this seems to have a negative impact on
the constraining power of the SIA data.
For not too large values of $p_T$, data obtained in $pp$ collisions are
in turn mainly sensitive to $D_g^{\eta}$ but in a limited
range of $z$, $0.4\lesssim z \lesssim 0.6$, as mentioned above.
Through the scale evolution, which couples quark and gluon FFs, these
data provide a constraint on $\xi^{\eta}_{u+\bar{u}}$.
In addition, the latest PHENIX data extend to a region of $p_T$ where
quark fragmentation becomes important as well. 
To illustrate this quantitatively, Fig.~\ref{fig:fractions} shows the relative fractions
of quarks and gluons fragmenting into the observed eta meson as a function
of $p_T$ in $pp$ collisions for PHENIX kinematics.
As can be seen, quark-to-eta FFs become dominant for $p_T\gtrsim 10\,\mathrm{GeV}$.

The $\chi^2$ profile for the truncated moment of the gluon, $\xi^{\eta}_g$, 
is the result of an interplay between the PHENIX run~'06 $pp$ data and the 
SIA data sets which constrain the moment $\xi^{\eta}_g$ towards smaller
and larger values, respectively. This highlights the complementarity of the
$pp$ and SIA data. SIA data have an impact on $\xi^{\eta}_g$ mainly through
the scale evolution in the energy range from LEP to BABAR. In addition,
SIA data provide information in the entire range of $z$, whereas the
$pp$ data constrain only the large $z$ part of the truncated moment $\xi^{\eta}_g$.
Consequently, the corresponding $\chi^2$ profile for $z_{\min}=0.4$ or $0.5$ 
would be much more dominated by $pp$ data.
In general, the other data sets from PHENIX \cite{ref:phenix2006} do not
have a significant impact on any of the truncated moments shown in 
Fig.~\ref{fig:profiles-ffs} due to their limited precision and covered
kinematic range.

Compared to pion and kaon FFs \cite{ref:dsspion}, all $\chi^2$ profiles in 
Fig.~\ref{fig:profiles-ffs} are significantly less parabolic, which prevents
one from using the Hessian method \cite{ref:hessian} for estimating 
uncertainties.
More importantly, the shapes of the $\chi^2$ profiles reflect 
the very limited experimental information presently available to extract
eta FFs for all flavors reliably. Another indication in that direction are the different
preferred minima for the values of the $\xi_i^{\eta}$ by the SIA and $pp$ data,
although tolerable within the large uncertainties.
Our fit is still partially driven by the set of
assumptions on the functional form of and relations among different FFs,
which we are forced to impose in order to keep the number of free fit 
parameters at level such that they can be actually determined by data.
Future measurements of eta production in SIA, $pp$ collisions,
and, in particular, SIDIS are clearly needed to test the assumptions made 
in our analysis and to further constrain the $D_i^{\eta}$.

The large variations found for the individual FFs in Fig.~\ref{fig:profiles-ffs}
are strongly correlated, and, therefore, their impact on uncertainty estimates 
might be significantly reduced for certain observables.
If, in addition, the observable of interest is only sensitive to a limited range
of hadron momentum fractions $z$, than the corresponding $\chi^2$ profile may
assume a more parabolic shape.

In order to illustrate this for a specific example, we compute 
the $\chi^2$ profiles related to variations in the theoretical estimates
of the single-inclusive production of eta mesons in $pp$ collisions at
PHENIX kinematics \cite{ref:phenix-run6}.
The results are shown in Fig.~\ref{fig:profiles-pp} for four different values
of $p_T$ along with the individual contributions to $\Delta \chi^2$ from the
SIA and $pp$ data sets.
As anticipated, we find a rather different picture as compared to Fig.~\ref{fig:profiles-ffs},
with variations only ranging from $5$ to $10\%$
depending on the $p_T$ value and tolerating $\Delta \chi^2/\chi^2=2\%$.
The corresponding uncertainty bands are also plotted in Fig.~\ref{fig:hadronic06} above
for both $\Delta \chi^2=1$ and $\Delta \chi^2/\chi^2=2 \%$
and have been obtained for the other $pp$ data from PHENIX \cite{ref:phenix2006}
shown in Figs.~\ref{fig:hadronic2g} and \ref{fig:hadronic3pi} as well.

The uncertainties for $pp \to \eta X$
are smallest for intermediate $p_T$ values, where the 
latest PHENIX measurement \cite{ref:phenix-run6} is
most precise and the three data sets \cite{ref:phenix2006,ref:phenix-run6} have maximum overlap,
and increase towards either end of the $p_T$ range of the run~'06 data.  
In particular at intermediate $p_T$ values, the main constraint comes from
the PHENIX run~'06 data, whereas SIA data become increasingly relevant at low $p_T$.
The previous $pp$ measurements from PHENIX \cite{ref:phenix2006} are limited to
$p_T\lesssim 11\,\mathrm{GeV}$ and have considerably larger uncertainties
and, hence, less impact on the fit.

\section{Conclusions\label{sec:conclusions}}
%
A first global QCD analysis of eta fragmentation functions at NLO accuracy has been
presented based on the world data from electron-positron
annihilation experiments and latest results from proton-proton collisions.
The obtained parameterizations \cite{ref:fortran} reproduce all data sets very well 
over a wide kinematic range.

Even though the constraints imposed on the eta meson fragmentation functions
by presently available data are significantly weaker than those for pions or kaons,
the availability of eta FFs extends the applicability of the pQCD framework 
to new observables of topical interest. Among them are the double-spin asymmetry for eta production
in longitudinally polarized proton-proton collisions at RHIC, eta meson
production at the LHC, possible medium modifications in the hadronization in the 
presence of a heavy nucleus, and predictions for future semi-inclusive lepton-nucleon scattering
experiments.

The obtained FFs still depend on certain assumptions, like $SU(3)$ symmetry
for the light quarks, dictated by the lack of data constraining the flavor separation
sufficiently well. Compared to FFs for other hadrons they show interesting patterns
of similarities and differences which can be further tested with future data.

\begin{acknowledgments}
%
We are grateful to David R.~Muller for help with the BABAR data.
CAA  gratefully acknowledges the support of the U.S.~Department of Energy 
for this work through the LANL/LDRD Program. The work of FE and JPS was 
supported by grants no.~DE-FG02-04ER41301 and no.~DE-FG02-94ER40818, respectively.
This work was supported in part by CONICET, ANPCyT, UBACyT, BMBF, and the 
Helmholtz Foundation.
\end{acknowledgments}



\begin{thebibliography}{99}
\bibitem{ref:fortran} A {\sc Fortran} package containing our NLO
set of eta FFs can be obtained upon request from the
authors.
\end{thebibliography}
\end{document}